\documentclass[12pt]{iopart}
\usepackage{iopams}  
\usepackage{graphicx}
\newcommand{\APJ}{{\it Astrophys. J.\/} }
\newcommand{\MNRAS}{{\it Mon. Not. Roy. Astronom. S.\/} }

\begin{document}

\title{Constraints on Quintessence Using Recent Cosmological Data}

\author{L P L Colombo and M Gervasi}

\address{Dipartimento di Fisica ``G Occhialini'', Universit\`a di
Milano-Bicocca, Piazza della Scienza, 3 I20126 Milano, Italy and INFN
Sezione di Milano-Bicocca} \ead{loris.colombo@mib.infn.it}

\begin{abstract}
Recent data, including the three--year WMAP data, the full 2dF galaxy
power spectrum and the first--year data of the Supernova Legacy
Survey, are used to constrain model parameters in quintessence
cosmologies. In particular, we discuss the inverse power--law (RP) and
SUGRA potentials and compare parameter constraints with those for
$\Lambda$CDM. Both potentials fit current observations with a goodness
of fit comparable or better than $\Lambda$CDM.  The constraints on the
energy scale $\Lambda_{\rm DE}$ appearing in both potential expressions
are however different. For RP, only energy scales around the
cosmological constant limit are allowed, making the allowed models
quite similar to $\Lambda$CDM.  For SUGRA, $\Lambda_{\rm DE}$ values
approximately up to Electroweak energy scale are still allowed, while other
parameter intervals are slightly but significantly displaced. In
particular a value of the primeval spectral index $n_s = 1$ is still
allowed at the 95$\, \%$ c.l., and this can have an impact on
constraints on possible inflationary potentials.
\end{abstract} 


\submitto{JCAP}

\maketitle

\section{Introduction}
\label{sec:intro}

In the last two decades, observations of high redshift
supernovae~\cite{sn} showed that the expansion of the
Universe is undergoing a phase of positive acceleration. Together with
measurements of Large Scale Structure~\cite{lss} and of anisotropies
of the Cosmic Microwave Background (CMB) temperature and
polarization~\cite{cmbold}, these data can be explained if about 70 \%
of the energy density of our universe is made up by a component with
negative pressure, dubbed Dark Energy (DE).

The basic candidate for DE is the cosmological constant, $\Lambda$,
with an equation of state $w \equiv p/\rho = -1$. As showed by the
three-year Wilkinson Microwave Anisotropy Probe (WMAP)
observations~\cite{wmap3:meas}, a power--law $\Lambda$CDM model with
adiabatic initial conditions is a good fit to most cosmological
data~\cite{wmap3:cosmo}. Within the framework of a $\Lambda$CDM
cosmology, the combination of the three-year WMAP (WMAP3) observations with
LSS and SNIa data, allowed to measure several fundamental cosmological
parameters with percent accuracy. However, a cosmological constant has
severe problems on theoretical grounds, in particular fine--tuning and
cosmic coincidence.
 
Several alternatives to a cosmological constants have been proposed,
including a perfect fluid with constant $w \ne -1$;
self--interacting scalar fields, with a time depending $w > -1$
(quintessence~\cite{quint,bm}) or $w
< -1$ (phantom~\cite{rob:phantom}), or with non--canonical kinetic
term (k--essence~\cite{kess}; quintom~\cite{quintom});
modifications of General Relativity~\cite{gr:mod}.

Quintessence models are particularly interesting due to the existence
of {\em tracker} potentials. Under the proper assumptions for the
potential~\cite{track}, the energy density of quintessence follows
closely the evolution of the dominant component, though decreasing
more slowly. In addition, the scalar field evolution does not depend
on the initial conditions, lessening fine--tuning and related
problems. On the contrary, dynamics depend on the current energy
density of DE and on the shape of the scalar field potential $V(\phi)$
and its parameters, which should be added to the usual cosmological
ones during data analysis.

Several authors used a combination of WMAP3 data and
other cosmological data, to study DE properties. The WMAP team
placed constraints on the value of $w$, assumed constant, while other
works studied the evolution of $\rho_{\rm DE}$~\cite{wm06}, or assumed
phenomenological parameterization for $w(a)$~\cite{zhao06}.  However, the
resulting constraints may depend on the choice of parameterization
(e.g.~\cite{jassal}) and, in general, do not translate
directly into constraints on the shape of the potential.

In this work we use recent cosmological data to directly constrain the
shape of two classes of quintessence potentials: the Ratra--Peebles
Potential (RP~\cite{rp}) and the SUGRA potential~\cite{bm}. The plane
of the paper is as follow. In Sec.~\ref{sec:models} we outline the
models studied together with the data and the method used. In
Sec.~\ref{sec:results} we describe our results and discuss them in
Sec.~\ref{sec:discussion}.

\section{Models and Data}
\label{sec:models}
In the limit of spatial homogeneity, the equation of motion for a
quintessence scalar field $\phi$ is given by
\begin{equation}
\ddot{\phi} +2 H \dot{\phi} + a^2 V'(\phi) = 0
\label{eq:motion}
\end{equation}
where dots denote derivatives with respect to conformal time $\tau$, the
Hubble parameter $H \equiv \dot{a}/a$, and we used natural units, $ c = \hbar
=1$. The energy and pressure associated with the field are given by
\begin{equation}
\rho_\phi = {1 \over 2} \left( {\dot {\phi} \over a} \right)^2 + V(\phi)~; 
~~~~~~~ p_\phi = {1 \over 2} \left( {\dot{\phi} \over a}\right)^2 - V(\phi)~. 
\label{eq:prho}
\end{equation}
Together with Friedman equations,~\eref{eq:motion} and ~\eref{eq:prho}
allow to solve for the background evolution in a quintessence
cosmology, once the the potential $V(\phi)$, and the density
parameters of the different components $\Omega_{\rm b}$, $\Omega_{\rm
c}$, etc., are assigned.  At variance with a cosmological constant,
quintessence is not a smooth component and perturbations in the scalar
field need to be accounted for~\cite{weller03}. In the Synchronous
gauge they obey the equation
\begin{equation}
\ddot{\varphi} +2 H \dot{\varphi} +a^2 V''(\phi) + {1\over 2}
\dot\mathfrak{h}\dot{\phi} + \nabla^2 \varphi = 0~,
\label{eq:pert}
\end{equation}
where $\varphi \equiv \delta \phi$ and $\mathfrak{h}$ is
the trace part of the metric perturbation. Perturbation evolution is
also fixed by the potential. 

Quintessence cosmologies, then, clearly depend on shape of $V(\phi)$,
which in principle is totally arbitrary. Considering only tracker
potentials allow to ease the fine tuning problem of $\Lambda$CDM
cosmologies.  As said in Sec.~\ref{sec:intro}, in this work we focus
on two classes of tracker potentials, RP and SUGRA:
\begin{eqnarray}
V(\phi) = {\Lambda_{\rm DE}^{4+\alpha} \over \phi^\alpha} ~, & \qquad
\qquad {\rm RP} \\ 
V(\phi) = {\Lambda_{\rm DE}^{4+\alpha} \over
\phi^\alpha} \, \exp \left( 4\pi \phi^2 \over m_p^2 \right)~, & \qquad
\qquad {\rm SUGRA}
\label{eq:su}
\end{eqnarray}
here $m_p$ is the Planck mass. The former was one of the first tracker
potentials studied, and is characterized by a slowly varying
equation of state. The latter originates within the context of
Supergravity theories and the corresponding equation of state
undergoes a rapid transition when the field becomes dominant. Both
potentials depend on two parameters, the slope $\alpha$ and the energy
scale $\Lambda_{\rm DE}$. For a given choice of parameters, the scalar
field associated with the two potentials evolve in a similar way until
the epoch of DE domination; at that point in a SUGRA model $w(a)$
rapidly falls toward values $-1 \lesssim w(1) \lesssim -0.9$, while in
a RP model $w(a)$ decreases more slowly, and stays at considerably
higher values.

In a cosmological context, $\alpha$ and $\Lambda_{\rm DE}$ are not
independent; fixing the current energy density of DE, $\rho_{\rm DE,
0}$, leaves only one free parameter. Notice that for $\alpha \simeq
0$, we recover the behaviour of the cosmological constant.  Here, we
take $\Lambda_{\rm DE}$ as a free parameter; however, different
conventions are possible~\cite{rp:wmap1,calvo}.

In this work, we test RP and SUGRA models using recent data, and try
to place constraints on the value of $\Lambda_{\rm DE}$. We assume a
power--law spectrum of primordial density fluctuations with adiabatic
initial conditions,and restrict our analysis to flat cosmologies with
no massive neutrinos, so that $\Omega_{\rm DE} = 1 - \Omega_{\rm b} -
\Omega_{\rm c}$. The models studied are thus specified by seven parameters:
the physical baryon, $\omega_{\rm b} \equiv \Omega_{\rm b} h^2$, and CDM,
$\omega_c$, densities; the angular size of the sound horizon at
recombination $\theta_s$; the slope, $n_s$, and amplitude, $A_s$, of
the power--spectrum of density fluctuations; the optical depth to
reionization, $\tau$; and $\lambda \equiv \rm{Log}_{10} (\Lambda_{\rm
DE} /\rm{GeV})$. Notice that $h$ is the reduced Hubble parameter,
${ H_0} = 100 h $Km/s/Mpc. This choice of parameters is particularly
efficient in minimizing the impact of the degeneracies of CMB angular
power spectra on data analysis~\cite{degen}. To explore the parameter
space, we used a modified version of the COSMOMC package. We assumed a
flat priors on all parameters; in particular $ -11.75 \lesssim \lambda
\lesssim 19 $. The upper limit corresponds to the Planck mass, while
the lower one is set by $(\Lambda_{\rm DE})^4 \simeq 10^{-47}
\rm{GeV}^4$, i.e. the cosmological constant energy density in a
concordance $\Lambda$CDM model. We also tested that a different choice of
priors, in particular using $ H_0$ instead of $\theta_s$, does not
significantly alter our results.

Models were tested against  WMAP3 data, the galaxy power spectrum
$P(k)$ of the full 2dF Galaxy Redshift Survey~\cite{2df:pdk} and the
distance measurements to high-redshift supernovae by the Supernova
Legacy Survey (SNSL~\cite{sn:astier}). Assuming the three sets of data
to be independent, the total likelihood is the product of the
individual likelihoods
\begin{equation}
\cal{L}{\rm tot} = \cal{L}_{\rm CMB} \times \cal{L}_{\rm LSS} \times
\cal{L}_{\rm SN}~.
\end{equation}
We also define an effective $\chi^2 \equiv -2 \ln (\cal{L})$. 

In the analysis of the CMB anisotropy spectrum, the WMAP team included
the contribution of secondary anisotropies due to Sunayev--Zel'dovich
(SZ) effect~\cite{wmap3:cosmo}. The SZ contribution was parametrized
in term of the amplitude of the signal, $A_{\rm SZ}$, relative to a
reference $\Lambda$CDM model and final constraints were obtained after
marginalization over $A_{\rm SZ}$. The effects of SZ marginalization
on $\Lambda$CDM estimates are very small. In addition, the SZ
contribution depends on the mass function and its evolution; these, in
turn, depend on the quintessence model considered and require
extensive numerical simulations~\cite{de:simul}. Thus, we set $A_{\rm SZ}
= 0$ here. We also did not include small scale data in our analysis
of CMB anisotropies. High--$\ell$'s mostly probe the recombination
history and the running of the spectral index, but are not significantly
affected by the quintessence models we considered.

Analysis of LSS data requires the introduction of an additional
parameter, $b$, accounting for the bias in the clustering of galaxies
with respect to the clustering of the matter field. Here, we assume
that $b$ is independent of scale in the range of $0.02~h\,{\rm
Mpc}^{-1} < k < 0.15~h\,{\rm Mpc}^{-1}$ considered~\cite{2df:pdk}, and
the code performs an analytic marginalization over it. In these range
of wavenumbers non--linear corrections are small~\cite{2df:pdk}, and
we do not include them in the analysis. For both SUGRA and RP
potentials we run 10 chains of $\sim 50000$ points each. We also run a
set of chains for a $\Lambda$CDM model. Comparing results for
quintessence with those for $\Lambda$CDM allows then to factor out
possible biases introduced by our choice of datasets, priors and
methods.

\section{Results}
\label{sec:results}

\Tref{tab:sugra} and \tref{tab:rp} summarizes our main results for
SUGRA and RP cosmologies, respectively. For each parameter $x$, except
$\lambda$, we list the mean value
\begin{equation}
\langle x \rangle = \int dx' {\cal L} (x') x'
\end{equation}
and the limits of the 68\% confidence interval, defined as appropriate
quantiles of the marginalized distribution, for different combinations
of data sets. For $\lambda$ we list the upper 95\% confidence limit.
\Tref{tab:lcdm}, instead, shows results for a basic 6--parameters
$\Lambda$CDM cosmology, in order to outline the effects of
quintessence.

\begin{figure}
\begin{center}
\includegraphics*[width=8cm]{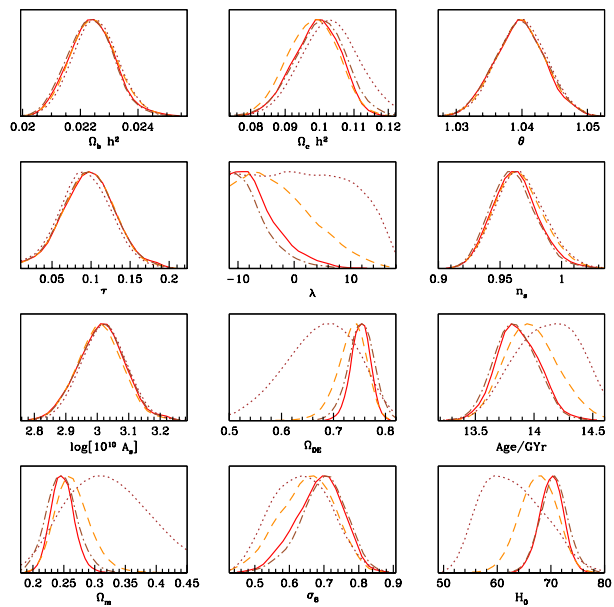}
\end{center}
\caption{\label{fig:su1d} Marginalized 1D constraints on basic and
derived parameters of SUGRA cosmology. Dotted lines show constrains from
WMAP3 only, dashed lines from WMAP3 + 2dF, dot-dashed from WMAP3 +
SNLS and solid lines from the combination of all data. WMAP3--only
constraints are very loose due to degeneracies between $\lambda$ and
other parameters, especially $\Omega_{\rm m}$.}
\end{figure}

\begin{figure}
\begin{center}
\includegraphics*[width=14cm,height=5cm]{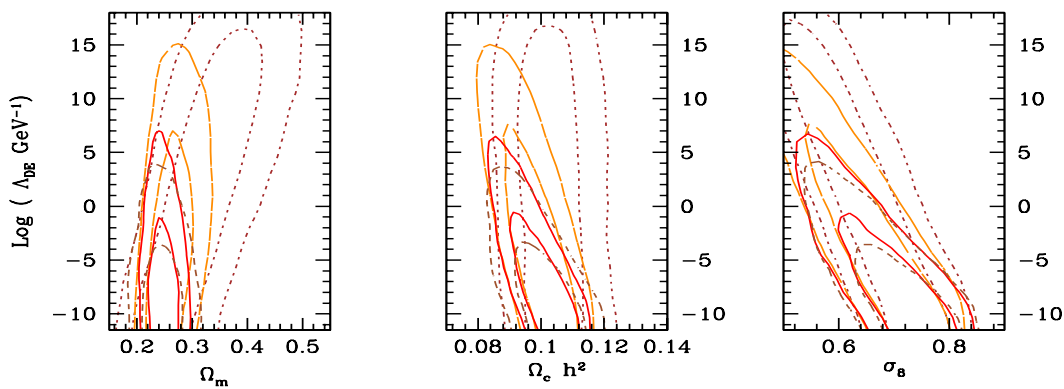}
\end{center}
\caption{\label{fig:su2d} Left: Joint 2D constraints in the ${ H_0}
  - \lambda$ plane for a SUGRA potential. Lines references are the
  same as~\fref{fig:su1d}. Notice the strong
  degeneracy between the two parameters when using only CMB
  data. Marginalization over $\lambda$ yields a high value of
  $\Omega_{\rm m}$ with a large uncertainty.  Adding additional data
  strongly suppress the high value of $\lambda$ and marginalized
  constraints are similar to results for $\Lambda$CDM.
  Middle: Joint constraints in the $\omega_c -\lambda$ plane. Right:
  Joint constraints in the $\sigma_8 -\lambda$ plane.}
\end{figure}

\begin{figure}
\begin{center}
\includegraphics*[width=8cm]{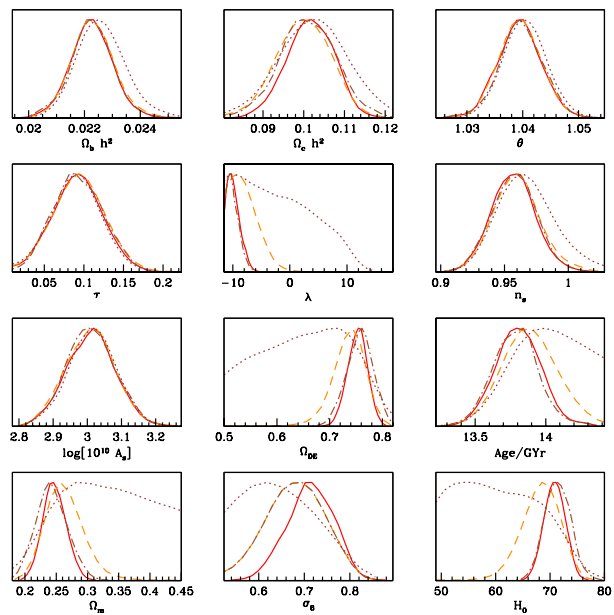}
\end{center}
\caption{\label{fig:rp1d} Marginalized 1D constraints on basic and
derived parameters of RP cosmology,  for the same combinations of
data sets as in~\fref{fig:su1d}.}
\end{figure}

\begin{figure}
\begin{center}
\includegraphics*[width=14cm,height=5cm]{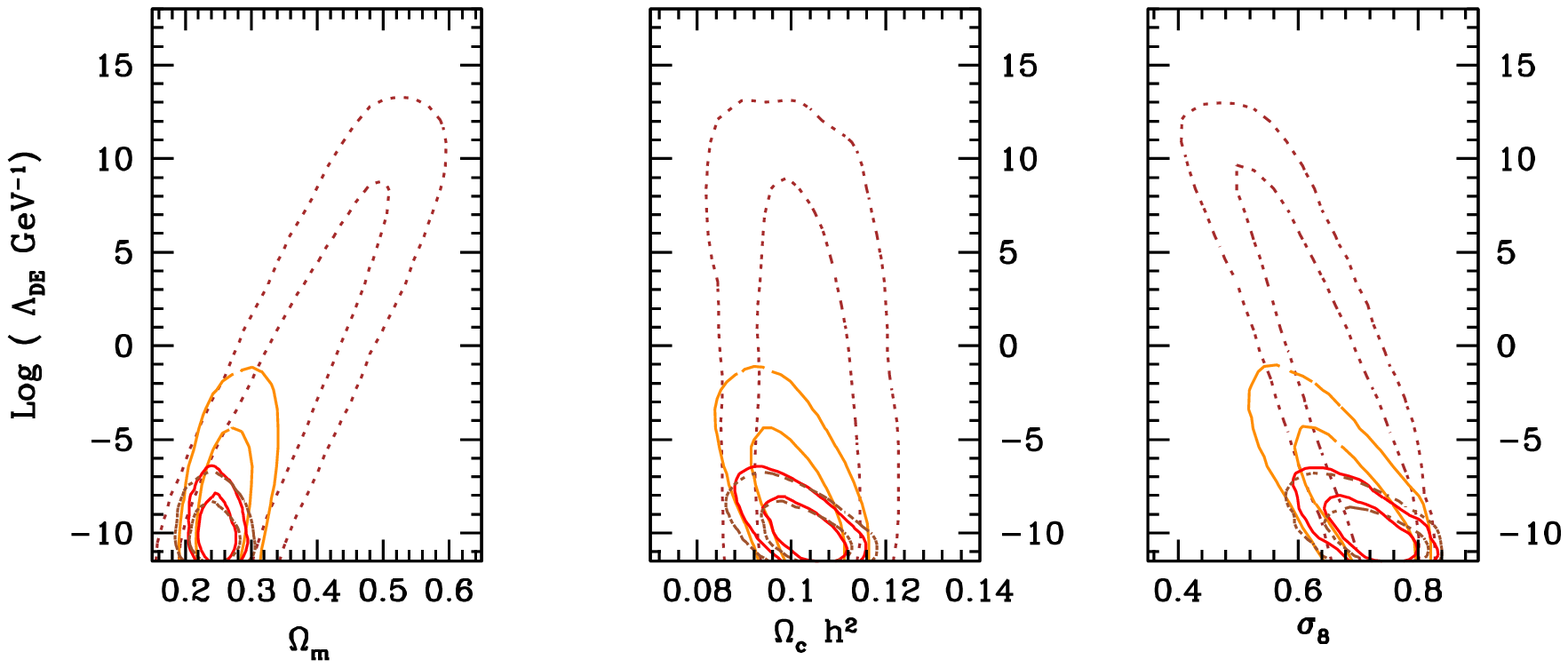}
\end{center}
\caption{\label{fig:rp2d} Left: Joint 2D constraints in the
  $\Omega_{\rm m} - \lambda$ plane for a RP potential. Lines
  references are the same as \fref{fig:su1d}.}
\end{figure}

In \fref{fig:su1d} and \fref{fig:rp1d}, we plot the marginalized
1D likelihoods, for some basic and derived parameters. Different
colours and lines types refer to different combinations of data (see
caption for details). In \fref{fig:su2d} and \fref{fig:rp2d} we
show the joint 2D confidence limits between $\Lambda_{\rm DE}$ and
other relevant parameters.

\begin{table}
\caption{SUGRA parameters. For all parameters except $\Lambda_{\rm DE}$
we show the mean value $\langle x \rangle = \int dx' {\cal L} (x') x'$
and corresponding 68\% confidence level interval, for the different
combinations of datasets used. For $\Lambda_{\rm DE}$ we show the 95\%
upper confidence limit.}
\label{tab:sugra}
\vglue0.2truecm
\begin{center}
\begin{tabular}{ccccc}
\hline
\rule[-1ex]{0pt}{3.5ex}
Parameter   &  WMAP3  &  WMAP3 + 2dF &  WMAP3 + SNLS   &  WMAP3 + 2dF  \\
            &         &              &                 & +SNLS \\
\hline
$100 \Omega_b h^2$  & $2.252^{+0.077}_{-0.078}$ 
                    & $2.246^{+0.078}_{-0.078}$  
                    & $2.234^{+0.074}_{-0.079}$  
                    & $2.242^{+0.073}_{-0.077}$  \\
$\Omega_c h^2$  & $0.1032^{+0.0082}_{-0.0082}$   
                & $0.0980^{+0.0072}_{-0.0074}$   
                & $0.1001^{+0.0071}_{-0.0075}$      
                & $0.0991^{+0.0069}_{-0.0069}$    \\
$ H_0  $        & $63.0^{+6.2}_{-5.8} $    
                & $67.6^{+3.4}_{-3.3} $    
                & $70.6^{+2.3}_{-2.5} $    
                & $70.1^{+2.1}_{-2.1} $     \\
$ \tau $        & $0.093^{+0.014}_{-0.015} $     
                & $0.098^{+0.015}_{-0.015} $     
                & $0.097^{+0.015}_{-0.015} $          
                & $0.098^{+0.015}_{-0.015} $       \\
$ n_s$          & $0.967^{+0.020}_{-0.020} $   
                & $0.965^{+0.019}_{-0.019} $   
                & $0.961^{+0.018}_{-0.018} $      
                & $0.963^{+0.017}_{-0.018} $       \\
${\rm Log}_{10} (\Lambda_{\rm DE})$   &  $< 14.9 $    
                                      &  $< 9.3$  
                                      &  $< 0.1$  
                                      &  $< 2.1$   \\
$\Omega_{\rm m}$  & $0.325^{+0.070}_{-0.069} $    
                  & $0.266^{+0.027}_{-0.026} $    
                  & $0.247^{+0.024}_{-0.024} $    
                  & $0.248^{+0.017}_{-0.018} $      \\
$\sigma_8 $       & $0.634^{+0.085}_{-0.086} $    
                  & $0.648^{+0.079}_{-0.082} $    
                  & $0.696^{+0.065}_{-0.063} $    
                  & $0.685^{+0.066}_{-0.069} $      \\

\hline
\end{tabular}
\end{center}
\end{table}

\begin{table}
\caption{RP parameters.}
\label{tab:rp}
\vglue0.2truecm
\begin{center}
\begin{tabular}{ccccc}
\hline
\rule[-1ex]{0pt}{3.5ex}
Parameter   &  WMAP3  &  WMAP3 + 2dF &  WMAP3 + SNLS   &  WMAP3 + 2dF  \\
            &         &              &                 & +SNLS \\
\hline
$100 \Omega_b h^2$  &$2.262^{+0.083}_{-0.083}$ 
                    &$2.232^{+0.072}_{-0.074}$ 
                    &$2.225^{+0.074}_{-0.070}$ 
                    &$2.225^{+0.070}_{-0.071}$ \\
$\Omega_c h^2$  & $0.1025^{+0.0083}_{-0.0083}$
                & $0.0999^{+0.0066}_{-0.0065}$ 
                & $0.1015^{+0.0067}_{-0.0067}$ 
                & $0.1016^{+0.0057}_{-0.0056}$   \\
$ H_0  $        & $ 59.7^{+8.7}_{-8.7} $
                & $ 68.1^{+3.3}_{-3.2} $
                & $ 71.3^{+2.3}_{-2.3} $
                & $ 70.8^{+1.9}_{-1.9} $     \\
$ \tau $        & $0.091^{+0.014}_{-0.014}$
                & $0.094^{+0.014}_{-0.015}$
                & $0.094^{+0.014}_{-0.014}$     
                & $0.092^{+0.014}_{-0.015}$     \\
$ n_s$          & $0.969^{+0.022}_{-0.023} $   
                & $0.960^{+0.017}_{-0.017} $   
                & $0.958^{+0.017}_{-0.016} $   
                & $0.957^{+0.016}_{-0.016} $     \\

${\rm Log}_{10} (\Lambda_{\rm DE})$   &  $<  8.9$    
                                      &  $< -3.6$  
                                      &  $< -7.9$  
                                      &  $< -7.7$   \\
$\Omega_{\rm m}$ & $0.369^{+0.109}_{-0.103} $    
                 & $0.266^{+0.025}_{-0.027} $     
                 & $0.245^{+0.025}_{-0.024} $     
                 & $0.247^{+0.019}_{-0.018} $         \\
$\sigma_8 $      & $0.600^{+0.080}_{-0.086} $
                  & $0.676^{+0.066}_{-0.063} $
                  & $0.712^{+0.057}_{-0.053} $
                  & $0.713^{+0.050}_{-0.050} $    \\

\hline
\end{tabular}
\end{center}
\end{table}

\begin{table}
\caption{$\Lambda$CDM parameters.}
\label{tab:lcdm}
\vglue0.2truecm
\begin{center}
\begin{tabular}{ccccc}
\hline
\rule[-1ex]{0pt}{3.5ex}
Parameter   &  WMAP3  &  WMAP3 + 2dF &  WMAP3 + SNLS   &  WMAP3 + 2dF  \\
            &         &              &                 & +SNLS \\
\hline
$100 \Omega_b h^2$  & $2.220^{+0.073}_{-0.072}$ 
                    & $2.220^{+0.072}_{-0.070}$ 
                    & $2.223^{+0.072}_{-0.072}$                    
                    & $2.219^{+0.071}_{-0.070}$    \\
$\Omega_c h^2$      & $0.1054^{+0.0079}_{-0.0080}$  
                    & $0.1070^{+0.0050}_{-0.0050}$  
                    & $0.1075^{+0.0058}_{-0.0058}$                      
                    & $0.1077^{+0.0044}_{-0.0044}$  \\
$ H_0  $            & $ 73.0^{+3.2}_{-3.2}  $       
                    & $ 72.2^{+2.0}_{-2.1}  $       
                    & $ 72.2^{+2.3}_{-2.3}  $       
                    & $ 71.9^{+1.8}_{-1.8}  $   \\
$ \tau $            & $0.090^{+0.013}_{-0.014} $    
                    & $0.086^{+0.013}_{-0.013} $    
                    & $0.087^{+0.013}_{-0.013} $                        
                    & $0.085^{+0.013}_{-0.012} $    \\
$ n_s$              & $0.955^{+0.016}_{-0.016} $   
                    & $0.953^{+0.016}_{-0.016} $   
                    & $0.953^{+0.016}_{-0.016} $                       
                    & $0.953^{+0.015}_{-0.016} $    \\
$\Omega_{\rm m}$    & $0.242^{+0.034}_{-0.035} $    
                    & $0.248^{+0.022}_{-0.022} $     
                    & $0.250^{+0.025}_{-0.025} $                        
                    & $0.252^{+0.018}_{-0.018} $     \\
$\sigma_8 $         & $0.758^{+0.048}_{-0.048} $                    
                    & $0.767^{+0.037}_{-0.037} $                    
                    & $0.770^{+0.041}_{-0.041} $                    
                    & $0.770^{+0.036}_{-0.036} $    \\

\hline
\end{tabular}
\end{center}
\end{table}

\section{Discussion}
\label{sec:discussion}

Using the first--year WMAP (WMAP1) data, several
authors~\cite{rp:wmap1} placed constraints on the slope of RP
potential, finding $\alpha \lesssim 2$. This constraint does not
directly translate into a constraint on the energy scale $\Lambda_{\rm
DE}$, as the $\alpha$--$\Lambda_{\rm DE}$ relation depends on
$\rho_{\rm DE,0}$. However, it was clear that allowed values fall well
below the $1-10^6$GeV range, which includes the Electroweak (EW) and
(possibly) the Supersymmetry (SUSY) breaking scales. Let us outline
soon that WMAP3 data do not improve this situation. While a SUGRA
potential still provides a fair fit of data for a physically relevant
range of $\Lambda_{\rm DE}$ values, this is not so for RP. We
therefore confirm that SUGRA is still a viable cosmology; on the
contrary, while RP is compatible with data, the cosmology it
yields has just minor differences from $\Lambda$CDM. Let us then
discuss first the SUGRA case, which leads to relevant shifts of some
parameters in respect to $\Lambda$CDM.

\subsection{SUGRA potential}
\label{ssec:sugra}
A first point to be outlined is that there is no appreciable
difference between the likelihood of SUGRA and $\Lambda$CDM, as the
$\chi^2$ value, for SUGRA, is only very marginally improved.

When only WMAP3 data are considered, the range of allowed
$\Lambda_{DE}$ values is essentially unconstrained: $\lambda =
\log_{10} (\Lambda_{DE}/{\rm GeV}) \lesssim 15$, at the 95\%
confidence level. Considering the whole set of data, the limit is
significantly more stringent: $\lambda \lesssim 2.1$ at
95\% c.l. Assuming for the other parameters their expectation
values, $\lambda = 2.1 $ corresponds to $\alpha = 3.36$. The
strong difference from RP is due to the fact that, at late times, when
$\phi$ approaches the Planck mass, the exponential term dominates the
potential, smoothing the dependence on the energy scale.

As far as other parameters are concerned, mean values and confidence
intervals are different from $\Lambda$CDM; shifts are more relevant
for $n_s$, $\Omega_{\rm m}$ and $H_0$. The shift of the primeval
spectral index $n_s$, in particular, affects the impact of WMAP3 data
on the nature of inflation ~\cite{wmap3:cosmo,inflation}. Using the
whole data set, we find for $\Lambda$CDM:
\begin{equation}
\nonumber
n_s = 0.953^{\,+0.015\,+0.031}_{\,-0.016\,-0.030}~,
\end{equation}
while for SUGRA:
\begin{equation}
\nonumber
n_s = 0.963^{\,+0.017\,+0.038}_{\,-0.018\,-0.032}~.
\end{equation}
Here we have quoted the mean, the 68\% and 95\% c.l. errors; the 95\%
errors include the contribution from the 68\% uncertainty. These
figures show that a Zel'dovich spectrum, excluded at the 95\%
c.l. ($n_s < 0.984$), if $\Lambda$CDM models are considered, is
compatible with the whole data set ($n_s < 1.001$), at the same
confidence level, for SUGRA cosmologies (see also~\cite{deinfl}).

More in detail, let us outline that, when using only WMAP3 data,
constraints on $\Omega_{\rm m}$, $H_0$ and $\sigma_8$ are very poor,
in qualitative agreement with results for WMAP1 data~\cite{noi:dual}.
In particular, $H_0 = 63 \pm 7 $Km/s/Mpc, $\Omega_{\rm m} = 0.32 \pm
0.07$ and $\sigma_8 =0.63 \pm 0.09$. This is due to a strong
degeneracy between $\lambda$ and $\Omega_{\rm m}$, which in turn
affects estimates of $H_0$ and $\sigma_8$, as shown
in~\fref{fig:su2d}. This degeneracy is qualitatively similar to the
geometric $w$--$\Omega_{\rm m}$ degeneracy in models with a constant
equation of state. Fixing all other parameters, the effective value of
$\langle w \rangle = \int \, da \, w(a) \Omega_{\rm DE}(a) /\int\, da
\, \Omega_{\rm DE}(a) $, is an increasing function of $\Lambda_{\rm
DE}$ and the field becomes dominant at earlier times than in
$\Lambda$CDM. When fitting CMB data, this behaviour yields higher $\Omega_{\rm
m}$ and lower $\sigma_8$ as $\Lambda_{\rm DE}$ increases. Notice that
there is no one--to--one relation $\Lambda_{\rm DE}$--$w(a)$, and
results for models with constant $w$ cannot be straightforwardly
applied to constrain the energy scale.

The main difference between WMAP1 and WMAP3 constraints is in the
marginalized distribution of $n_s$ and $\tau$. The Integrated
Sachs--Wolfe effect acts on low--$\ell$ anisotropy multipoles more
strongly in SUGRA than in $\Lambda$CDM. In WMAP1, this increases the
$\tau$--$n_s$ degeneracy, and both $\tau$ and $n_s$ are quite higher
for SUGRA than for $\Lambda$CDM. WMAP3 estimates of TE and EE spectra
allows a firm determination of $\tau$, and considering dynamical DE
does not significantly worsen constraints on $\tau$ and $n_s$. 

Including other datasets in the analysis strongly reduces the allowed
range for $\Lambda_{\rm DE}$, in turn significantly affecting the estimates of
parameters degenerate with the energy scale. Both LSS and SNIa data
exclude the highest values of $\Lambda_{\rm DE}$, corresponding to
models with an high effective $\langle w \rangle$. The
combination of WMAP3 and SNLS is particularly constraining for this
class of models, and adding also 2dF data has only minor effects
on estimates. In fact, current SN data seem to
prefer models with equation of state around or slightly lower than -1,
while for a quintessence model $ w(a) > -1$.

SUGRA also favours higher values of $\omega_{\rm b}$, although
differences are around 1/3 of the current uncertainty.  As discussed
above, $\Lambda_{\rm DE}$ affects estimates of $\tau$ and $n_s$ which
in turn affect $\omega_{\rm b}$ through the $\tau-n_s-\omega_{\rm
b}-A_s$ degeneracy. However, if additional data allow to establish a
clear difference in estimates for SUGRA and $\Lambda$CDM, Big Bang
Nucleosynthesis measures could allow to distinguish the two models.

Let us also outline that SUGRA models show an improvement in the
goodness of fit, when LSS data are included in the analysis. Such
improvement is modest, $\Delta \chi^2 \simeq -1$, and does not provide
a compelling evidence in favour of SUGRA cosmology. However, the
dynamical range of SUGRA models is still very wide and compatible with
the energy scales of fundamental physics, and at present they offer a
valid alternative to $\Lambda$CDM.

\subsection{RP potential}
\label{ssec:su}

Considering only WMAP3 data, results for RP are qualitatively similar
to SUGRA. Upper limits on $\Lambda_{\rm DE}$ are moderately tighter
than in SUGRA, $\lambda \lesssim 8.9$ (95\% c.l.), and discrepancies
between RP and $\Lambda$CDM estimates are greater that those between
SUGRA and $\Lambda$CDM.

However, adding other data strongly suppress the high tail of the
distribution on $\Lambda_{\rm DE}$; using all data, we find $\lambda
\lesssim -7.7$ (95\% c.l.), well below the relevant range for EW and
SUSY breaking.  Moreover, estimates for other parameters are very
similar to $\Lambda$CDM values, except for $\Omega_{\rm c} h^2$ and
$\sigma_8$. Both are lower than the corresponding $\Lambda$CDM figures
by about 1 standard deviation.

As already stated, previous works~\cite{rp:wmap1,calvo} used various
combinations of data to constrain the slope of RP potential, finding
that $\alpha \lesssim 2$. A direct comparison of the marginalized
distribution on $\alpha$ and $\Lambda_{\rm DE}$ is not immediate;
however, we notice that assuming $\Lambda_{\rm DE} = 10^{-7} {\rm
GeV}$ we find that $0.7 \lesssim \alpha \lesssim 0.8$ when
$\Omega_{\rm DE}$ varies in the range $0.1 - 0.9$ and $H_0 = 70
$Km/s/Mpc. For these values of the parameters, the range of initial
conditions for which $\phi$ reaches the tracking regime is very small,
and RP models have fine tuning problems similar to
$\Lambda$CDM~\cite{bludman}. Choosing initial conditions so that $\phi$
does not reach the tracking regime by the present time, could alter
the bounds on the potential parameters~\cite{notrack}. However,
without tracking, quintessence models lose one of their conceptual
advantages over other DE candidates.

Thus RP models, while still compatible with present data, do not seem
to be a strong alternative to a standard $\Lambda$CDM cosmology.

\subsection{Summary}
\label{ssec:summary}

We have used recent cosmological data to constrain the parameters of
two relevant quintessence potentials, Ratra--Peebles (RP) and SUGRA.
Both potentials depend on an energy scale, $\Lambda_{\rm DE}$, and the
corresponding cosmologies reduce to $\Lambda$CDM for $(\Lambda_{\rm
DE})^4 \simeq 10^{-47} {\rm GeV}^4$. 

Both for RP and SUGRA, current data allow to put only upper limits on
$\Lambda_{\rm DE}$, due to a strong degeneracy between $\Lambda_{\rm
DE}$ and $\Omega_{\rm m}$. When considering only the third--year WMAP
data, the degeneracy results in wider errors on several parameters,
including $\Omega_{\rm m}$, $H_0$ and $\sigma_8$. When LSS and,
especially, SN data are added to the analysis, constraints on
$\Lambda_{\rm DE}$ strongly tighten. For RP potential, only values
close to the cosmological limit are still allowed, and the model does
not offer clear advantages to $\Lambda$CDM.

For SUGRA potential, instead, values of $\Lambda_{\rm DE} \simeq 100$
GeV, close to weak interaction energy scale, are still compatible with
data. In addition, estimates of several parameters are different
from the corresponding $\Lambda$CDM values. In particular, the
spectral index, $n_s$, is compatible with a Zel'dovich spectrum at
less than 2 standard deviations. This could have a strong impact on
constraints on the shape of the inflation potential. SUGRA models,
thus, are still a viable and significant alternative to $\Lambda$CDM.

\ack We thank Roberto Mainini and Silvio Bonometto for helpful
discussions.

\end{document}